\documentclass[11pt,twocolumn,notitlepage,a4wide]{report}
\usepackage{graphicx}
\usepackage{epsfig}
\usepackage{fleqn}
\usepackage{amsmath}
\usepackage{natbib}
\usepackage{color}
\usepackage{setspace}
\usepackage{afterpage}
\usepackage[labelsep=period,font=small,labelfont=bf]{caption}

\newcommand{\matr}[1]{\boldsymbol{#1}}
\newcommand{\vect}[1]{\boldsymbol{#1}}

\setcitestyle{super,comma,open={},close={}}
\graphicspath{{./figures/}}

\title{GPEC, A REAL-TIME CAPABLE TOKAMAK EQUILIBRIUM CODE}
\author{M.\ Rampp$^1$,
R.\ Preuss$^2$,
R.\ Fischer$^2$
and the ASDEX Upgrade Team$^2$
\\
\medskip\\
$^{1}$ Max Planck Computing and Data Facility\\
Gie{\ss}enbachstra{\ss}e~2, 85748 Garching, Germany\\
\smallskip\\
$^2$ Max-Planck-Institut f\"ur Plasmaphysik,\\
Boltzmannstra{\ss}e\ 2, 85748 Garching, Germany\\
}

\begin{document}

\twocolumn[\begin{@twocolumnfalse}
\maketitle

%
\begin{abstract}
{\bf 
\noindent A new parallel equilibrium reconstruction code for tokamak plasmas
is presented. GPEC allows to compute equilibrium flux distributions
sufficiently accurate to derive parameters for plasma control 
within 1~ms of runtime which enables real-time applications at the
ASDEX Upgrade experiment (AUG) and other machines with a control cycle
of at least this size. 
The underlying algorithms are based on the well-established 
offline-analysis code CLISTE, following the classical concept of
iteratively solving the Grad-Shafranov equation and feeding in diagnostic
signals from the experiment.
The new code adopts a hybrid parallelization scheme for
computing the equilibrium flux distribution and extends the 
fast, shared-memory-parallel Poisson solver which we have 
described previously by a
distributed computation of the individual Poisson problems
corresponding to different basis functions. The code is based 
entirely on open-source software components and runs on standard 
server hardware and software environments.
The real-time capability of GPEC is demonstrated by performing 
an offline-computation of a sequence of 1000 flux distributions which are
taken from one second of operation of a typical AUG discharge and deriving the relevant 
control parameters with a time resolution of a millisecond. On current
server hardware the new code allows employing a grid size of $32
\times 64$ zones for the spatial discretization and up to 
15 basis functions. It takes into account about 90 diagnostic 
signals while using up to 4
equilibrium iterations and computing more than 20 plasma-control
parameters, including the computationally expensive safety-factor $q$ on
at least 4 different levels of the normalized flux.
}
\end{abstract}
\end{@twocolumnfalse}]
%
%
\clearpage
\newpage
\section{INTRODUCTION}\label{sect:introduction}

Reconstruction of the plasma equilibrium shape is a key
requirement for the operation of current and forthcoming tokamak experiments
\cite[e.g.][]{gormezano07,gribov07,mgg05,gia2013}. 
The most commonly
employed method numerically reconstructs the plasma equilibrium by
iteratively solving the two-dimensional Grad-Shafranov equation \cite{sha57,lus57} 
for the poloidal flux function ($\psi$), and using diagnostic
signals from the experiment as constraints \cite{lfg90,cliste}. 
In short, the source term of the Grad-Shafranov equation is expanded into a linear
combination of basis functions which allows formulating a linear regression problem for the expansion coefficients,
using the actual diagnostic signals and their forward-modelling based
on the numerical solution for $\psi$.
The problem can be tackled numerically by solving a number of
independent Grad-Shafranov-type equations --- one for each 
basis function of the expansion --- and employing a Picard-iteration  
procedure that allows evaluating the source term 
using the solution from the previous iteration step.

A variety of equilibrium-reconstruction codes based on this strategy \cite{cliste}
or similar approaches \cite{lao85,lfg90,blum12} exist and have been routinely used for
diagnostics and data analysis at the various tokamak experiments.
The codes EFIT \cite{lao85,lfg90} and CLISTE
\cite{cliste}, for example, have been employed at
DIII-D, EAST and ASDEX Upgrade (AUG), respectively,
a class of medium-sized tokamaks which are characterized by a control cycle on the
order of a millisecond. 
Until recently, however, computing times in the millisecond range 
were inaccessible to such "first-principles" codes, and hence their
applicability for real-time diagnostic analysis and plasma-control is 
very limited, unless the numerical resolution or the number of 
iteration steps is drastically reduced or other simplifying
assumptions like function parametrization \cite{braams86} are adopted \cite[e.g.][]{giannone10}.
Lately, three codes based on fast Grad-Shafranov solvers have been
able to demonstrate real-time capability: P-EFIT \cite{P-EFIT}, a GPU-based variant of the EFIT 
equilibrium-reconstruction code \cite{lao85,lfg90}, is able to compute one
iteration within 0.22~ms on a $65\times 65$ grid with 3 basis
functions. With a similar resolution the JANET 
software  \cite{Barp20122112,gia2013} has reached runtimes of
about 0.5~ms per iteration on a few CPU cores within the
LabVIEW-based system at AUG.
Using a single CPU core overclocked at 5 GHz, the LIUQE 
code \cite{Moret20151} which is deployed at the Swiss TCV experiment,
achieves a cycle time of about 0.2~ms with 
a spatial resolution of $28\times 65$ points. 
In all these cases, however, only a single iteration is performed
and only a very small number of basis functions can be afforded for the parametrization
of the source term of the Grad-Shafranov equation. While such
restrictions may well
be justifiable under specific circumstances \cite{fer98} the motivation for our work is to 
substantially 
push the limits of numerical accuracy in terms of spatial
resolution, number of basis functions, and number of iterations and 
thus the quality of the equilibrium solution. This is expected to
aid the generality and robustness of the application, in
particular with respect to variations in the plasma conditions.

In addition to the reliability and accuracy of real-time equilibrium
reconstructions, the sustainability and maintainability of the
codes is considered an important aspect.
Typical implementations of analysis tools in fusion science have a
rather long life time which often exceeds life times of commercial
hardware and software product cycles.
The long usability required for analysis tools poses special requirements
on the software chosen and the maintenance capabilities necessary to
adapt to changing environments.
Open-source software together with off-the-shelf computer hardware is
thought of being perfectly suited for these demands \cite{yonekawa12}. 
This in particular applies to all non-standardized components like, e.g.,
the implementation of complex, tailored, and evolving numerical algorithms for
equilibrium reconstruction. Commercial, closed-source 
software (like, e.g., compilers or numerical libraries) and 
hardware components may well comply with such demands, provided that 
they can be modularly interchanged using standardized interfaces.

To this end we have implemented a new, parallel equilibrium-reconstruc\-tion code, 
GPEC (Garching Parallel Equilibrium Code), which builds on the fast,
shared-memory-parallel Grad-Shafranov solver
we have developed previously \cite{gpec12} and a  
two-level hybrid parallelization scheme which was pointed out in the same paper.
The basic numerical model and functionality of the new code originate from
the well-established and validated algorithms implemented by the
equilibrium codes CLISTE \cite{cliste} and specifically its descendant IDE
\cite{fis13}, both of which are being used for comprehensive offline diagnostics and data 
analysis at the ASDEX Upgrade experiment. 
GPEC is based entirely on open-source software components, runs on standard 
server hardware and uses the same code base and computer architecture 
as employed for performing offline analyses with the IDE code. 
Such a strategy is 
considered a major advantage, in particular concerning verification
and validation 
of the code \cite{lfg90} and also its evolution: On the one hand, algorithmic and 
functional innovations in the offline physics modeling can gradually be taken over by 
the real-time version. On the other hand, high-resolution offline
analysis can directly take advantage of optimizations achieved for the real-time variant.

The new code enables computing the equilibrium flux distribution
and the derived diagnostics and control parameters within 1~ms of runtime, given a grid size of $32 \times 64$
zones with up to 15 spline basis functions for the discretization, and about 90
diagnostic signals. With a runtime of less than 0.2~ms for a single equilibrium iteration 
and about 0.2~ms required for computing a variety of more than 20 relevant control
parameters, one can afford 4
iterations for converging the equilibrium solution. The latter was checked to be sufficient for reaching accuracies of better than a percent for the relevant quantities.
As a proof of principle we shall demonstrate the real time capability 
of the new code by performing an offline analysis using data
from an AUG discharge.
To our knowledge the new GPEC code to date is one of the fastest (at a given numerical
accuracy) and most accurate (at a given runtime constraint) of its kind. 

\medskip

The paper is organized as follows: in Section~\ref{sect:algorithm} we
recall the basic equations and describe our new, hybrid-parallel 
implementation of the equilibrium solver and its
verification and validation. Its computational
performance and in particular real-time capability is demonstrated in 
Section~\ref{sect:application} on an example
application using real AUG data. Section~\ref{sect:conclusions} 
provides a summary and conclusions.

\section{ALGORITHMS AND IMPLEMENTATION}\label{sect:algorithm}

\subsection{Equilibrium Reconstruction}\label{sect:algorithm_equil}

\paragraph{Basic equations}
The Grad-Shafranov equation \cite{sha57,lus57} describing ideal magneto-hydrodynamic
equilibrium in two-dimensional tokamak geometry reads
\begin{equation}
\Delta^* \psi(r,z) = - 2 \pi \mu_0 r\, j_\phi\quad,
\label{gs}
\end{equation}
with cylindrical coordinates ($r$, $z$), the elliptic differential operator 
\begin{equation}
\Delta^* :=
r \frac{\partial}{\partial r} \frac{1}{r} \frac{\partial}{\partial r}
+
\frac{\partial^2}{\partial z^2}\quad,
\label{gsdiffop}
\end{equation}
and the poloidal flux function (in units of
Vs), $\psi(r,z)$.
The toroidal current density profile, 
\begin{equation}
j_\phi := 2 \pi \left(
   r \frac{\partial p (r, \psi) }{\partial \psi} 
 + \frac{F (\psi) }{\mu_0 r} \frac{{\rm d}F (\psi) }{{\rm d}\psi}\right)
,
\label{gs_current}
\end{equation}
consists of two terms, where $p(r, \psi)\equiv p(\psi)$ is the plasma pressure (isotropic case) and
$F(\psi) = r B_\phi = (\mu_0/2 \pi) I_{\rm pol}$ is
proportional to the total poloidal current, $I_{\rm pol}$.

\paragraph{Numerical solution}

The classical strategy \cite{cliste,cliste12,fer98} for numerically
solving Eq.\ (\ref{gs})
is based on an expansion of the current density profile $j_\phi$
on the right-hand-side into a linear combination 
of a number $N=N_p+N_F$ of basis functions,
\begin{equation}
\begin{split}
j_\phi&(r,z)  = \\ & 2 \pi \left(r \sum_{k=1}^{N_p} c_k \pi_k (\psi) + \frac{1}{\mu_0\,r}\sum_{k=1}^{N_F} c_{N_p+k} \varphi_k (\psi)\right).
\end{split}
\label{rhsdecomp}
\end{equation}
In each cycle of a Picard-iteration scheme, a number of $N$ Poisson-type equations,
\begin{equation}
\Delta^* \psi_k  = - 4 \pi^2 \mu_0 r^2 \pi_k (\psi) \quad (k=1\dotsc N_p),
\label{gsphi1}
\end{equation}
and
\begin{equation}
\Delta^* \psi_{N_p+k}  = - 4 \pi^2 \varphi_k (\psi)       \quad (k=1\dotsc N_F),
\label{gsphi2}
\end{equation}
are solved individually, where the solution $\psi$ from the last iteration 
step is used for evaluating the right-hand side.
The updated flux distribution is then given by
\begin{equation}
\psi = \sum_{k=1}^{N} c_k \psi_k
\quad,
\label{bfdecomp}
\end{equation}
which follows from Eqs.~(\ref{gs},\ref{rhsdecomp}) and the linearity 
of the operator $\Delta^*$.  
The unknown coefficients $c_k$ are determined 
by experimental data: Using the $N$ distributions $\psi_k$ obtained
from Eqs.~(\ref{gsphi1},\ref{gsphi2}) the so-called response matrix
$\matr{B}:=\{b_l(\psi_k)\}_{l=1,\dotsc,M,k=1,\dotsc N}$, consisting of predictions for
a set of $M$ diagnostic signals $\{m_l\}_{l=1\dotsc,M}$  is calculated.
The prediction $b_l(\psi)$ is some (linear) function employing the flux
distribution $\psi$ to produce the forward-modelled signal.
All measurements are located within the poloidal flux grid such that the
poloidal flux outside the grid is not needed.
Linear regression of the response matrix with the measured
diagnostic signals finally yields the coefficients $c_k$ for
Eq.~(\ref{bfdecomp}). This procedure is iterated until certain convergence criteria are
fulfilled.

The private flux region close to the divertor is treated differently,
although the same poloidal flux as inside the plasma occurs. 
Since there are no measurements of the current distribution in the private flux
region and the chosen functional form for the current decay was proven
to be of minor importance for the equilibrium reconstruction, the
current is chosen to decay approximately exponentially with a decay
length of about 5 mm, starting at the last closed flux surface.

\bigskip

\noindent
The algorithm can be summarized as follows:
  \begin{enumerate}

  \item evaluate right-hand sides $g_k(r,z):=4 \pi^2
    \mu_0 r^2 \pi_k (\psi^{i-1})$ and $g_{N_p+k}(r,z):=4 \pi^2 \varphi_k (\psi^{i-1})$
    of Eqs.~(\ref{gsphi1},\ref{gsphi2}), using the solution, $\psi^{i-1}$, from the previous iteration step (or initialization). 
  \item solve the Poisson-type equations $\Delta^* \psi_k = - g_k(r,z)$
    individually for each $k=1,\dotsc,N$. An
    additional Poisson problem arises from a 
    convergence-stabilization procedure which introduces 
    the vertical plas\-ma position as an auxiliary free parameter \cite{cliste, Moret20151}.  
  \item evaluate $\{b_l(\psi_k)\}_{l=1,\dotsc,M, k=1,\dotsc,N}$ to
    construct the response matrix $\matr{B}$. 
    Additional columns $\{b_{l,N+1}\}_{l=1,\dotsc,M}$ and $\{b_{l,N+1\dots N+1+N_\mathrm{ext}}\}_{l=1,\dotsc,M}$ arise from
    the abovementioned convergence-stabilization procedure and from
    deviations in the currents measured at a number of $N_\mathrm{ext}$
    external field coils to account for wall shielding and plasma-induced
    wall currents, respectively. 
  \item perform linear regression on $\matr{B}\cdot \vect{c} = \vect{m}$ to obtain the set of
    coefficients $\{c_k\}_{k=1,\dotsc,N}$.  
  \item evaluate new solution $\psi^i= \sum_{k=1}^{N} c_k \psi_k + \psi^\mathrm{ext}$, 
  adding contributions from the measured and fitted deviating
currents in external coils, $\psi^\mathrm{ext}$.
  \item go back to step 1. until convergence is reached, e.g.\ in
    terms of the
    maximum norm evaluated over all grid points, ${\lVert\psi^i-\psi^{i-1}\lVert}_\infty$.
  \end{enumerate}

\noindent
The response matrix $\matr{B}$ is already prepared for additional measurements
of the motional Stark effect (MSE), the Faraday rotation, the pressure
profile (electron, ion and fast particle pressure), the q-profile 
(e.g. from MHD modes), the divertor tile currents constraining 
$F(\psi)$ on open flux surfaces, the measurements of loop voltage
and of iso-flux constraints \cite{fis13}.
These additional measurements are used in the IDE code in the off-line
mode, but are not subject of the present work due to the lack of reliable on-line
availability. 
The relatively large number of basis functions (and hence fit parameters), which
is motivated by the need to allow for equilibria sufficiently flexible to address all
occurring plasma scenarios, requires application of regularization
constraints. While in the IDE code additional smoothness (curvature
and amplitude) requirements are applied to the source profiles $p'$ and
$FF'$ (which effectively adds additional columns to the response matrix), GPEC
currently adopts a simpler ridge-regression procedure\cite{hoe70}.

\paragraph{Implementation and parallelization}

The implementation of GPEC starts out from the serial offline-analysis
code IDE \cite{fis13}, thus closely following the concepts of the
well-known CLISTE code \cite{cliste}. Specifically, for each of the
two sets of basis functions, 
$\{\pi_k\}_{k=1\dotsc N_P}$ and $\{\varphi_k\}_{k=1\dotsc N_F}$,
GPEC employs a basis of cubic spline polynomials which are defined
by a number of $N_p$ and $N_F$ knots at positions $x_i$, respectively, 
with $\pi_k(x_i)=\delta_{ki}$, $\varphi_k(x_i)=\delta_{ki}$ and natural boundary conditions.
Bicubic Hermite interpolation is employed for evaluating the 
flux distribution $\psi_k$ and its gradients (which are required for computing the
magnetic field) at the positions of the magnetic probes and 
the flux loops in order to calculate the corresponding 
forward-modelled signals for the response matrix.
Thus, besides achieving higher-order interpolation accuracy,
smoothness of the magnetic field is guaranteed by construction \cite{pre07}.

GPEC utilizes the fast, thread-parallel Poisson solver we have 
developed previously \cite{gpec12}. It is based on the two-step scheme
for solving the Poisson equation with Dirichlet boundary conditions
\cite{hal75} and
replaces the cyclic-reduction scheme which is traditionally employed in the 
 codes GEC and CLISTE \cite{lac76,cliste} by a parallelizable Fourier-analysis method for 
decoupling the linear system into tridiagonal blocks \cite{pre07}. For details
on the implementation and computational performance we refer to refs.~\cite{gpec,gpec12}.

As briefly sketched in ref.~\cite{gpec12} the main idea of the new code is to 
exploit an additional level of
parallelism by distributing the individual Poisson problems for
determining each of the $\psi_k$ to different MPI\footnote{The "Message
  Passing Interface" standard \cite{mpiforum_url}.}-processes. 
The algorithm, together with notes on the parallelization (relevant
MPI communication routines are noted in brackets) is
summarized as follows.

\medskip

Each process $P_{k=1,\dotsc,N}$ is assigned to a
single basis function\footnote{In general the 
implementation supports an even distribution of the $N+1$ Poisson
problems to a number of $N'$ processes, provided $N'$ divides $N+1$.}, and 
process $P_{N+1}$ is dedicated to the
additional Poisson problem that arises from the introduction of the 
vertical shift parameter \cite{cliste}. Accordingly, each process $P_k$:  
\begin{enumerate}
\item computes $g_k(r,z)$ using the solution $\psi(r,z)$ from the previous
  iteration step (or initialization) \\\null\hfill\emph{thread-parallelization over $(r,z)$-grid}
\item employs fast Poisson solver \cite{gpec12} to solve $\Delta^* \psi_k = - g_k(r,z)$ \\\null\hfill \emph{thread-parallelization} \cite{gpec12}
\item computes column $\vect{b}(\psi_k)$ of the response matrix
  \\\null\hfill\emph{thread-parallelization over $(r,z)$-grid} 
\item gathers columns $\vect{b}(\psi_{k\prime})$ computed by the other
  processes $P_{k\prime\ne k}$ and assembles response matrix $\matr{B}$
  \\\null\hfill\emph{collective communication} ({\tt MPI\_Allgather})
\item performs linear regression on $\vect{m}=\matr{B}\cdot
  \vect{c}$ \\\null\hfill\emph{thread-parallelization of linear algebra routines}
\item gathers all $c_{k\prime}\psi_{k\prime}$ computed by the other
  processes $P_{k\prime\ne k}$ and computes
  $\psi=\sum_{k\prime=1}^N c_{k\prime}\psi_{k\prime} $ \\\null\hfill\emph{collective
  communication} ({\tt MPI\_Allreduce})
\end{enumerate}

The distribution of the Poisson problems to different processes comes at the price of
collective communication (steps 4 and 6), in which all processes
combine their data with all others, using MPI routines 
{\tt MPI\_Allgather} and {\tt MPI\_Allreduce}, respectively.

\subsection{Computation of Plasma-Control Parameters}\label{sect:postprocessing}

Given the solution for the poloidal flux function $\psi(r,z)$,
GPEC computes more than 20 derived quantities which are
relevant for plasma control.
Two of them, the $(r,z)$-coordinates of the magnetic axis,
$R_\mathrm{mag}$ and $Z_\mathrm{mag}$ 
can be obtained immediately by searching the maximum of the
equilibrium flux.
For the rest of the quantities four major computational
tasks (labelled A--D in the following compilation) have to be carried
out. Their results allow evaluating the individual control parameters 
(summarized under the bullet points) in a numerically 
straightforward and inexpensive way.
For details on the physical definition and derivation of these
quantities, see, e.g.\ Ref.~\cite{stacey12}.
Implementation details are given further below.

\begin{enumerate}
\item[(A)] determination of grid points which are enclosed by the separatrix
\begin{itemize}
\item coordinates $(R_\mathrm{geo},Z_\mathrm{geo})$ of the geometric
axis, defined as the area-weighted integral within the separatrix
$\int (r,z) {\rm d}S/S$.
\item the horizontal and vertical minor plasma radius is given by
$a_{\rm hor}=2 \sqrt{ \int r^2 {\rm d}S/S - R_{\rm geo}^2 }$ and
$b_{\rm ver}=2 \sqrt{ \int z^2 {\rm d}S/S - Z_{\rm geo}^2 }$,
respectively, which determines the elongation $k=b_{\rm ver}/a_{\rm hor}$.
\end{itemize}
\item[(B)] identification of contour lines $(r(t),z(t))$ with
$\psi(r(t),z(t))=\psi_c$ for a number of particular values of the
normalized flux $\psi_c$
\begin{itemize}
\item from the separatrix curve (with $\psi_c$ taken as the value at
the innermost x-point which defines also the last-closed flux surface
LCFS) a number of geometric properties
can be derived straightforwardly such as
the ($r,z$)-coordinates of the uppermost and lowermost point (in
$z$-direction) on the plasma surface,
or 
$R_{\rm in}$ ($R_{\rm aus}$) as the $R$-coordinate of the
innermost (outermost) point on the plasma surface.
\item the length of the poloidal perimeter $l_{\rm pol}$ of the plasma
  is computed by integration over the LCFS.
\item the safety factor $q$ is computed by integration over contours
defined by levels $\psi_c\in [0.25,0.50,0.75,0.95]$.
\end{itemize}
\item[(C)] computation of the toroidal plasma current
$I_\mathrm{tor}=\int j_\phi {\rm d} S$ (Eq.~\ref{gs_current})
\begin{itemize}
\item the indicators for the current-center are given by the
following current-weighted integrals of the current density
$R_{\rm squad}= \sqrt{ \int r^2 j_\phi {\rm d} S / I_\mathrm{tor} }$,
$Z_{\rm squad}=        \int z   j_\phi {\rm d} S / I_\mathrm{tor}  $.
\end{itemize}
\item[(D)] computation of the pressure distribution $p(r,z)$ and the
  poloidal magnetic field
\begin{itemize}
\item the total energy content of the plasma is given by
$W_{\rm MHD} = 3/2 \int p {\rm d} V$.
\item the same integration can be used for computing the
poloidal beta parameter, $\beta_{\rm pol} = \int p {\rm d} V / Z_\beta$, 
employing the normalization constant 
$Z_\beta = \mu_0 \cdot V_{\rm plasma} \cdot I_{\rm tor}^2
  /(2 \cdot l_{\rm pol}^2)$ (as used in CLISTE) with $l_{\rm pol}$ and $I_{\rm tor}$
as specified under (B) and (C), respectively.
\item the plasma self-inductivity
$l_\mathrm{i} = (2 \mu_0 Z_\beta)^{-1} \int B_\mathrm{pol}^2 {\rm d} V$
is given by an integral over the squared poloidal magnetic field 
$B_\mathrm{pol}^2=(2\pi r)^{-2}\cdot((\partial_r \psi)^2+(\partial_z \psi)^2)$.
\item $dR_{\rm XP}$, the difference of the poloidal flux at the two
x-points divided by ($2\pi R_{\rm aus}$) and by the poloidal magnetic
field at ($R_{\rm aus},Z_{\rm mag}$): $( \psi_{\rm XP1} - \psi_{\rm
XP2}) / |\nabla \psi|_{(R_{\rm aus},Z_{\rm mag})}$.
\end{itemize}
\end{enumerate}

\paragraph{Implementation and parallelization}

The algorithms for computing the plasma-control
parameters and their basic implementation are taken directly 
from the IDE code. We emphasize
that unlike other real-time codes  \cite{Moret20151} 
GPEC makes no algorithmic simplifications compared with the offline variant of 
the code. In particular, GPEC uses the same spatial grid resolution 
as adopted for equilibrium reconstruction and employs the high-order 
interpolation schemes (cubic splines, bicubic Hermite interpolation) of IDE.
In the following we briefly summarize the main concepts adopted for the
computationally most expensive routines:

\begin{enumerate}
\item[(A)]
for performing the integrations $\int \dots {\rm d}S/S$, 
grid points located outside the separatrix are masked out
according to a comparison with the value of $\psi$ at the x-point.
\item[(B)]
a custom contour-finding algorithm is used which employs quadratic interpolation 
to obtain the values at the intersections of the spatial grid with the
contour curve.
Quantities on the LCFS, like $l_\mathrm{pol}$, are
determined by quadratic extrapolation from three equidistant 
contour levels close to the plasma surface.
\item[(C)] for the evaluation of the toroidal plasma current 
$I_{\rm tor}=\int j_\phi {\rm d} S$ the current density $j_\phi(r,z)$ is computed 
according to
Eq.~(\ref{rhsdecomp}). 
The summands $c_k \pi_k (\psi)$ and $c_{N_p+k} \varphi_k (\psi)$
in principle would be available 
from the last equilibrium iteration but are distributed among
processors (cf.\ Sect.~\ref{sect:algorithm_equil}). It turns out that locally
recomputing $\pi_k (\psi)$ and $\varphi_k (\psi)$ by using the converged 
equilibrium solution $\psi$ (a copy of which is available on
each processor) is faster than collecting the individual summands by MPI communication.
\item[(D)] the computation of the pressure distribution $p(r,z)$ 
is based on the identity $\partial_\psi p (\psi(r,z))=\sum_{k=1}^{N_p} c_k\pi_k (\psi)$ (cf.\ Eq.~\ref{rhsdecomp}). 
For fast evaluation of $p(r,z)$ on the poloidal mesh,
an equidistant grid of $\psi_l, l=1,\dotsc,12$ points is constructed.
For each $\psi_l$ the corresponding pressure
contribution $\pi_k(\psi_l)$ is computed by
numerical integration starting at the plasma boundary.
A cubic spline interpolation is used to 
evaluate $\pi_k(\psi(r,z))$ on this grid. The gradients for
computing the poloidal magnetic field are obtained using centered finite
differences of the neighboring grid points (consistent with the
derivatives used to define the bicubic Hermite interpolation
polynomials for intra-grid interpolation, cf.\ Sect.~\ref{sect:algorithm_equil}).
\end{enumerate}
As shall be shown in Sect.~\ref{sect:performance} the computing time
is dominated by finding and integrating along contours as well as by
evaluating the pressure distribution.
Accordingly, the computations are grouped into the following 
mutually independent tasks which can be assigned to different 
MPI tasks, with only a few scalar quantities to 
be communicated for evaluating the final results:
\begin{itemize}
\item computation of the pressure distribution $p(r,z)$ and the
  poloidal magnetic field
\item handling of 3 contour levels for extrapolation to the last closed flux surface
\item handling of 4 contour levels corresponding to the set of safety factors
  $q_{25}, q_{50}, q_{75}, q_{95}$
\item determination of the x-point and handling of 1 contour level
  corresponding to the separatrix curve
\end{itemize}
Within each MPI task, OpenMP threads are used,
e.g.\ for parallelizing the computation of the pressure over the 
individual points of the $(r,z)$-grid or for parallelizing over 
different contour levels.    

\subsection{Verification and Validation}
The codes CLISTE and IDE have been extensively validated and verified by 
application to ASDEX Upgrade data and by means of code comparisons. Thus, only 
those parts of the new GPEC code need to be verified which are treated
differently from the IDE code. The procedure is greatly facilitated by the 
fact that both codes implement the same algorithms and share the same code base. 
The only differences in the GPEC implementation with respect to IDE are
1) the prescription of a fixed number of equilibrium iterations and 2)
the use of only 12 instead of 30 integration points 
for computing the pressure distribution $p(r,z)$ from its gradient
$\partial_\psi p (\psi)$ (see item D above) on every grid point $(r,z)$.

\begin{figure*}[!ht]
  \centering
  \includegraphics[width=0.99\hsize,angle=0]{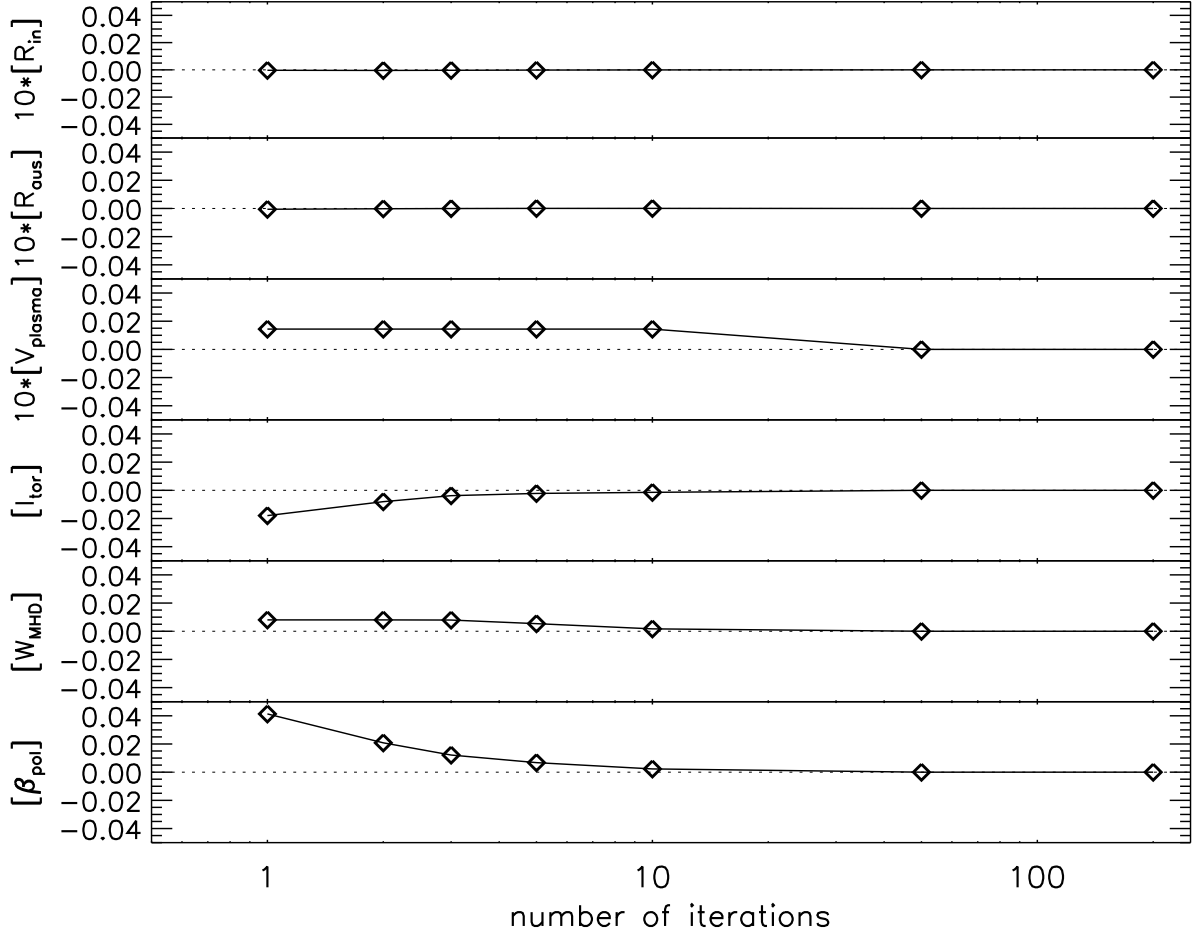}
  \caption{Sensitivity of selected parameters to the number of equilibrium iterations for
    computing $\psi(r,z)$. For each quantity the relative deviation from
    its converged value is shown as a function of the number of iterations. 
    Note that the deviations for the geometric quantities, $R_\mathrm{in}$,
    $R_\mathrm{aus}$, and the total volume of the plasma, $V_\mathrm{plasma}$ are multiplied by a factor of 10.  
    Data was taken from AUG shot \#23221 at time $t=3.015$~ms.}
  \label{fig:sens_iter}
\end{figure*}

Figure~\ref{fig:sens_iter} shows the sensitivity of a number of selected parameters
to the number of equilibrium iterations for an arbitrarily chosen time point 
($t=3.015$~ms) of AUG discharge \#23221.
By using the solution $\psi$ from the last time point as an initial value,
the plasma geometry (parameters $R_\mathrm{in}$, $R_\mathrm{aus}$, and the total 
volume of the plasma, $V_\mathrm{plasma}$) is already very well determined.
Within only a few iteration steps accuracies of better than a percent compared with the
converged solution (obtained with 200 iterations) are obtained also for 
the quantities which depend on the
pressure distribution ($\beta_{\rm pol}$, $W_{\rm MHD}$), or on the toroidal 
current ($I_{\rm tor}$). For the
real-time demonstration presented in the subsequent section we shall fix
the number of iterations to four and show that the accuracy level of 1~\% is
maintained over a time sequence of 1~s with 1000 time points.

\begin{figure*}[!th]
  \centering
  \includegraphics[width=0.99\hsize,angle=0]{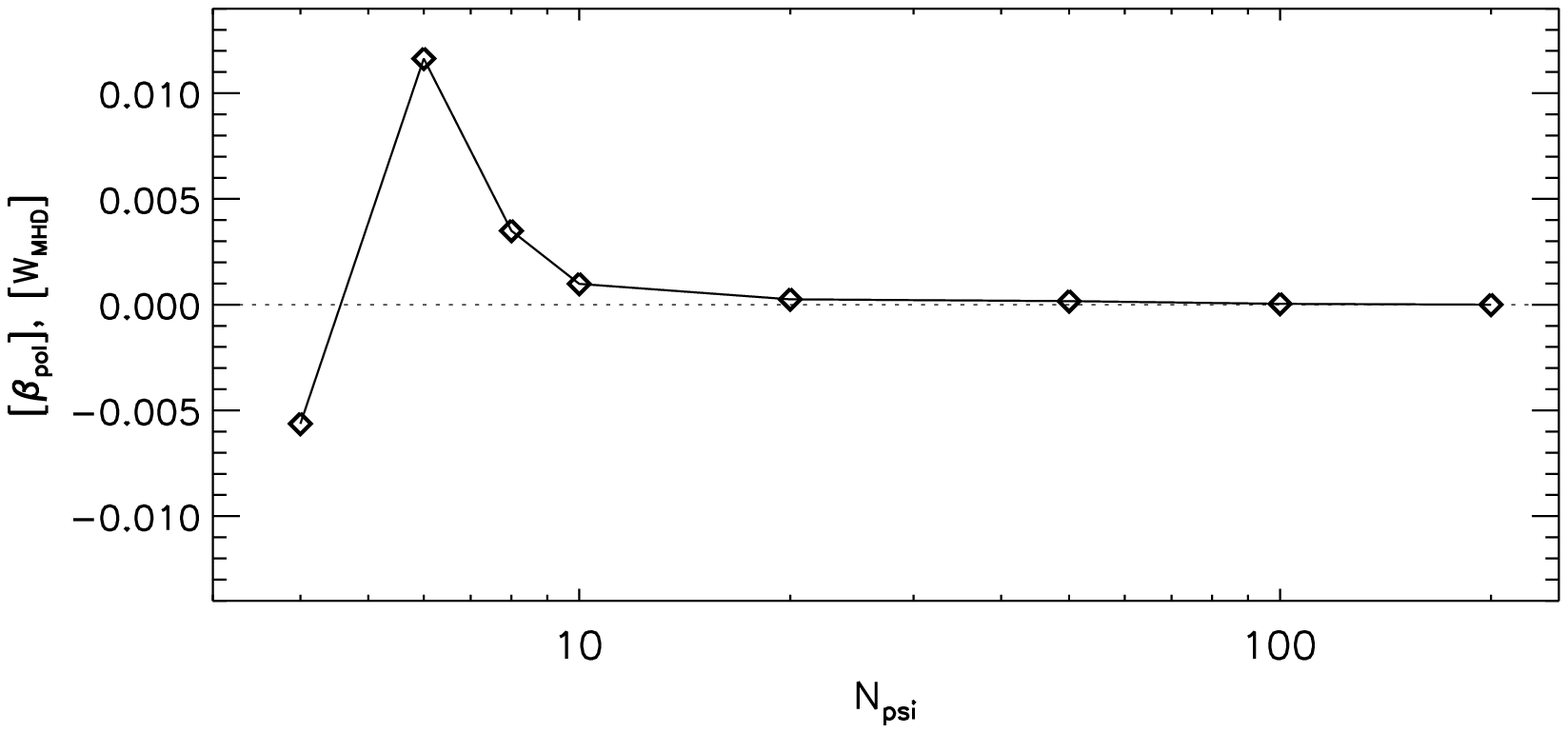}
  \caption{Sensitivity of $\beta_{\rm pol}$ and of the total energy content
    $W_{\rm MHD}$ to the number of integration
    points, $N_\mathrm{psi}$, spent for computing the pressure from 
    its gradient on every grid point $(r,z)$. The relative 
    deviation from the converged value is shown as a function of the number of integration
    points, $N_{\rm psi}$.
    Data was taken from AUG shot \#23221 at time $t=3.015$~ms.}
  \label{fig:sens_npsi}
\end{figure*}

Figure~\ref{fig:sens_npsi} shows that a number of 12 integration points for
the numerical integration of $\partial_\psi p (\psi)$ is more 
than sufficient in order to reach sub-percent accuracies for the control 
parameters $\beta_{\rm pol}$ and $W_{\rm MHD}$, both of which are
proportional to $\int p {\rm d} V$ (see item D above). 
The time for computing $p(r,z)$ is proportional to the number of integration points
and scales almost ideally with the number of threads which motivates our
specific choice of 12 points as a multiple of 6 threads that shall be employed 
for the majority of computations (cf.\ Sect.~\ref{sect:performance}).

\begin{figure*}[!h]
  \centering
  \vspace*{-3cm}
  \includegraphics[width=0.99\hsize,angle=0]{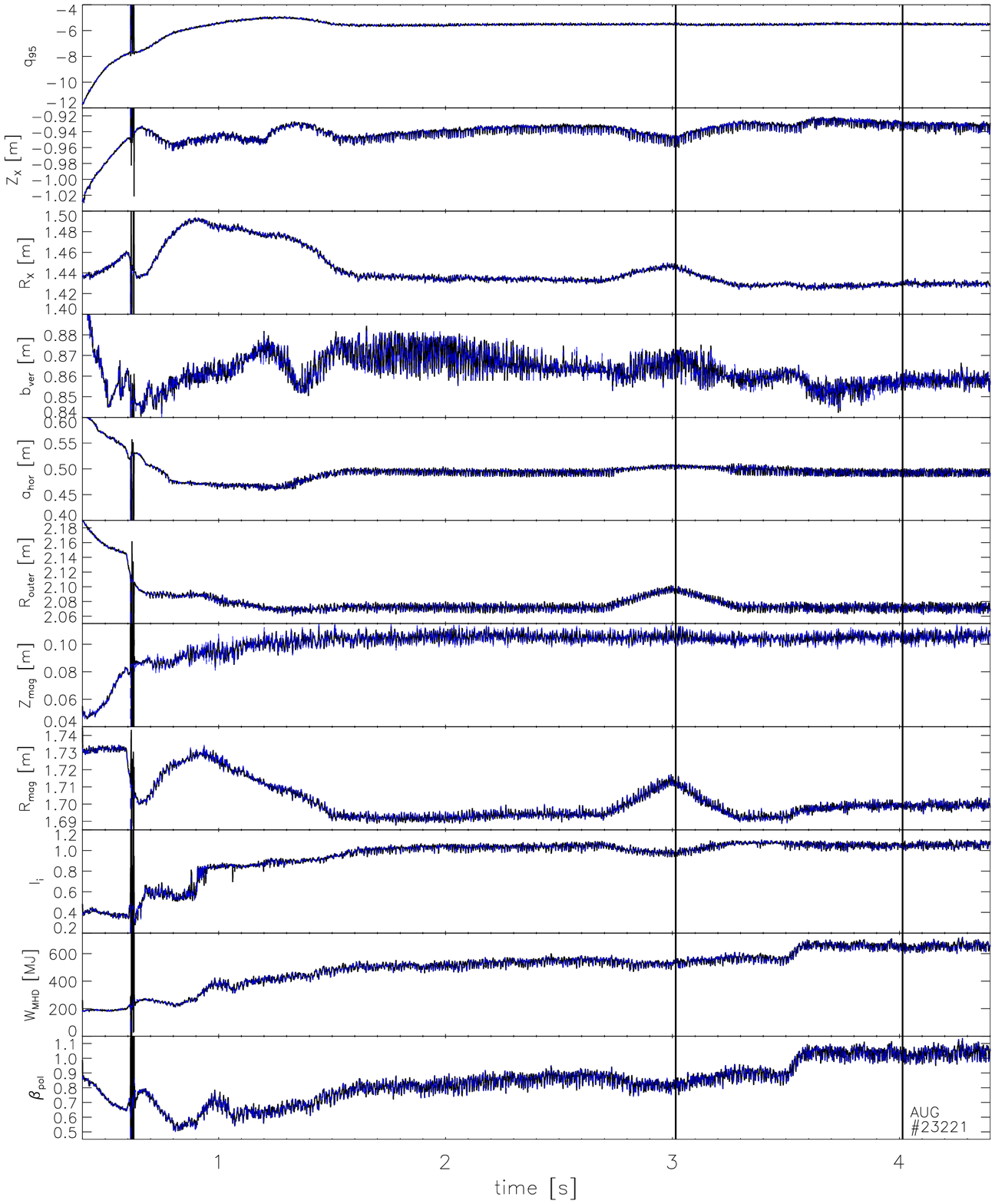}
  \caption{Time evolution of selected parameters for AUG shot
    \#23221 as computed with the new real-time code during an extended
    time window between $t=0.4~\mathrm{s}$ and $t=4.4~\mathrm{s}$. 
    The time interval $[3.014~\mathrm{s},4.014~\mathrm{s}]$ which is
    chosen for performance analysis
    is indicated by vertical lines. The individual plots show
    the coordinates $(R_\mathrm{mag},Z_\mathrm{mag})$ and 
    $(R_\mathrm{X},Z_\mathrm{X})$
    of the
    magnetic axis, and of the x-point, respectively, 
    the horizontal and vertical minor plasma radius,
    $a_\mathrm{hor}$ and $b_\mathrm{ver}$, the plasma
    self-inductivity, $l_\mathrm{i}$, the poloidal beta parameter,
    $\beta_\mathrm{pol}$, the total energy content of the plasma,
    $W_\mathrm{MHD}$, and the safety factor $q_{95}$ (see
    Sect.~\ref{sect:postprocessing} for the definition of the
    quantities). 
    The solid, black lines correspond to a "real-time" application
    which is restricted to 4 equilibrium iterations. For comparison,
    values derived from  converged equilibrium solutions (convergence
    criterion ${\lVert\psi^i-\psi^{i-1}\lVert}_\infty < 10^{-4}$) are
    shown as dashed, blue lines. A standard spatial 
    resolution of $(32,64)$ grid points and 7 basis functions were employed.}
  \label{fig:shot_23221alltime}
\end{figure*}

\section{APPLICATION PERFORMANCE}\label{sect:application}

\subsection{Demonstration of Real-Time Capability}

\begin{figure*}[!h]
  \centering
  \vspace*{-3cm}
  \includegraphics[width=0.99\hsize,angle=0]{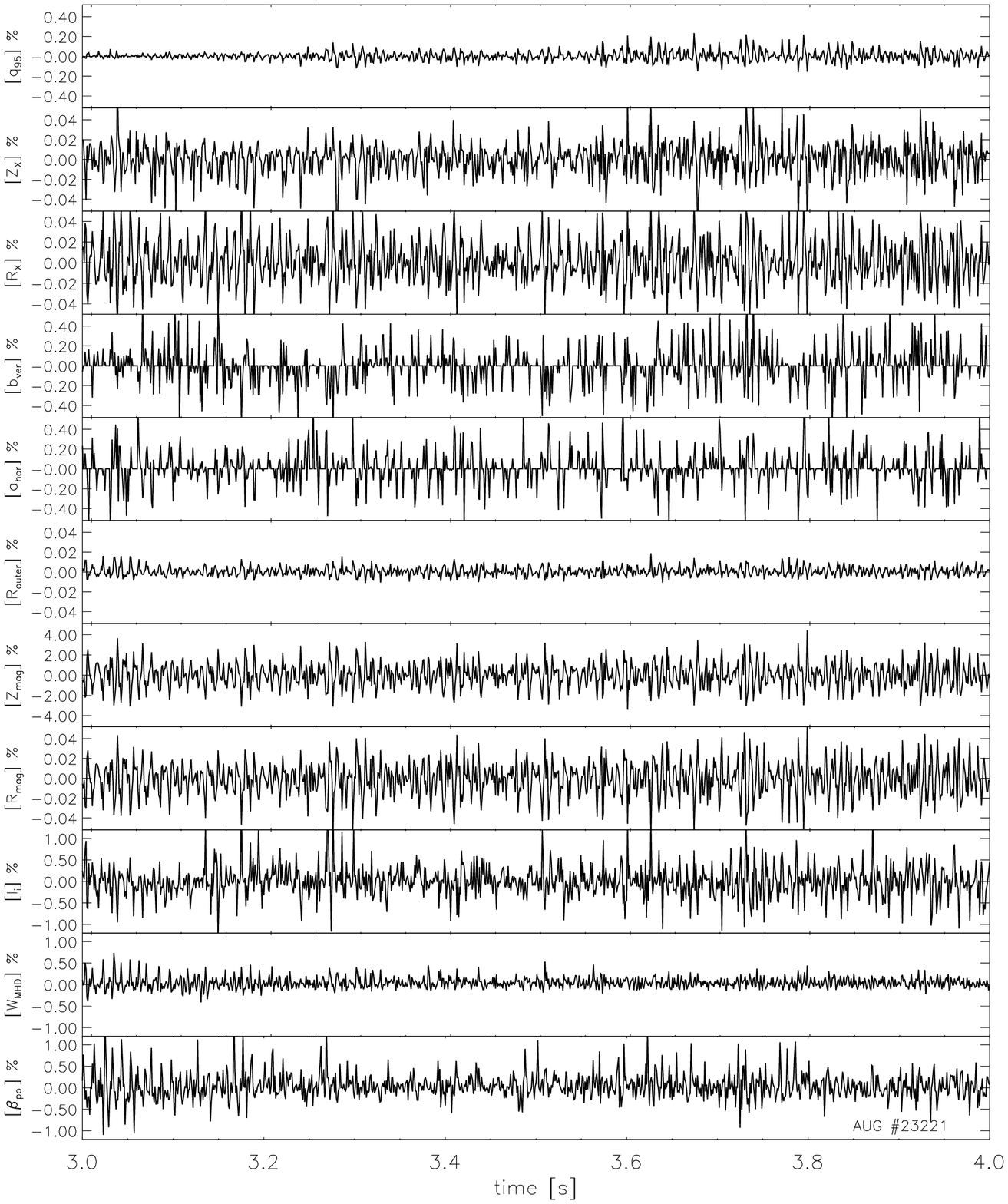}
  \caption{Evolution of the relative deviation (denoted by square brackets)
    of the quantities which were computed after 
    4 equilibrium iterations, from the values computed with converged
    equilibrium solutions (convergence
    criterion ${\lVert\psi^i-\psi^{i-1}\lVert}_\infty <
    10^{-4}$) during the time interval chosen for performance
    analysis (as indicated by vertical lines in
    Fig.~\ref{fig:shot_23221alltime}). A standard spatial 
    resolution of $(32,64)$ grid points and 7 basis functions were employed.}
  \label{fig:shot_23221-accuracy}
\end{figure*}

In order to demonstrate the real-time capability of the new code
a post-processing run is performed using data from a typical AUG
discharge. 
We chose AUG shot number \#23221, and focus on a time window 
between  $t_1=3.014~\mathrm{s}$ and $t_{1000}=4.013~\mathrm{s}$ with 
1000 time points, corresponding to a time resolution of exactly 1~ms. 
As shall be shown below this simulation of 1000 time points 
can be performed on standard server hardware within less than 
a second of computing time.


\begin{table*}[!ht]
\centering
\small
\begin{tabular}{|l|l|r|r|r|r|r|r|}
\hline
resolution     & $T_\mathrm{tot}$ & $T_\mathrm{eq}$ & $T_\mathrm{ctl}$ & cores & MPI & threads\\
$(N_r,N_z,N\!\!+\!\!1)$ & [ms] & [ms] & [ms] &(nodes) & tasks & per task\\ 
\hline
$(32,64,6)$   & 0.85 & 0.62 & 0.18 & 24 (1) &  6 & 4 \\
$(32,64,8)$   & 1.08 & 0.87 & 0.17 & 24 (1) &  4 & 6 \\
$(32,64,8)$   & 0.88 & 0.67 & 0.17 & 48 (2) &  8 & 6 \\
$(32,64,12)$  & 0.97 & 0.74 & 0.19 & 72 (3) & 12 & 6 \\
$(32,64,16)$  & 1.55 & 1.32 & 0.18 & 24 (1) &  4 & 6 \\
$(32,64,16)$  & 1.26 & 1.02 & 0.18 & 48 (2) &  8 & 6 \\
$(32,64,16)$  & 0.99 & 0.73 & 0.21 & 96 (4) & 16 & 6 \\
\hline                            
$(64,128,6)$  & 2.07 & 1.55 & 0.39 & 48 (2) &  6 & 8 \\
$(64,128,8)$  & 1.88 & 1.28 & 0.48 & 48 (2) &  8 & 6 \\
$(64,128,12)$ & 2.02 & 1.39 & 0.49 & 72 (3) & 12 & 6 \\
$(64,128,16)$ & 2.04 & 1.40 & 0.51 & 96 (4) & 16 & 6 \\
\hline

\end{tabular}
\caption{Overview of the total runtime $T_\mathrm{tot}$
  per time point ($2^\mathrm{nd}$ column) for different combinations
  of the numerical resolution ($1^\mathrm{st}$ column)
  and compute resources. The latter are given in terms of the total number of cores (nodes)
  ($5^\mathrm{th}$ column), of MPI tasks ($6^\mathrm{th}$ column) and
  of OpenMP threads per MPI task ($7^\mathrm{th}$ column). $T_\mathrm{tot}$ is the sum
  of the runtime $T_\mathrm{eq}$ ($3^\mathrm{rd}$ column) for computing the
  equilibrium distribution $\psi$ with 4 iteration steps (cf.\ Sect.~\ref{sect:algorithm_equil}), and $T_\mathrm{ctl}$ ($4^\mathrm{th}$ column) for
  deriving plasma-control parameters (cf.\ Sect.~\ref{sect:postprocessing}). The benchmarks were performed on Xeon E5-2680v3 "Haswell"
  CPUs (2.5 GHz, 24 cores per node) and used data from AUG shot \#23221.}
\label{tab:pide_timing}
\end{table*}

The GPEC code computes a 
fixed number of four equilibrium iteration steps for every time point and
uses the equilibrium solution $\psi(t_n)$ at time point 
$t_n$ as the initial value for the next time point $t_{n+1}$.
A standard spatial resolution of $(N_r,N_z)=(32,64)$ grid points is
  employed, and $N=7$ basis functions are used.
The response matrix $\matr{B}$ 
contains signals from 61 magnetic probes and 18 flux loops and 
in addition takes into account 10 external coils.
Figure~\ref{fig:shot_23221alltime} provides an overview of the time evolution of a number of
characteristic parameters, such as selected plasma-shape parameters, energy
content and safety factor $q_{95}$ (black, solid lines). For
reference, the figure shows an
extended time window between $t=0.4~\mathrm{s}$ and
$t=4.4~\mathrm{s}$, which includes
the dynamic start-up phase of the discharge with significant variations in the plasma
parameters, and also shows results from a run which converges the  
solution $\psi$ until the maximum changes on the grid, 
${\lVert\psi^i-\psi^{i-1}\lVert}_\infty$, are below $10^{-4}$ (blue, dashed lines). 
Compared with the converged reference run one notices good overall agreement
of the "real-time" simulation with a fixed number of 4 equilibrium
iterations throughout the extended time interval, even during the
initial start-up phase which lasts until approx.\ 1.5~s. It is
hence justified to confine the analysis on the --- arbitrarily chosen ---
time interval, $[3.014~\mathrm{s},4.013~\mathrm{s}]$ in the "plateau" phase of the discharge.
Focusing on this time window, Figure~\ref{fig:shot_23221-accuracy}
shows that the relative errors for most quantities are at smaller than one
percent. The comparably large variations of $Z_{\rm mag}$ are due to its small absolute
value. The absolute scatter of $Z_{\rm mag}$ is in the order of mm.
Note that the cycle time of $\Delta t=1$~ms is much shorter than the
timescale of the changes of plasma conditions (cf.\ the increase of
$W_\mathrm{MHD}$, $\beta_\mathrm{pol}$ at $t\approx 3.5$~s). Thus the 
equilibrium solution, even if not fully converged, can 
easily follow such secular trends, which is reflected by the fact 
that there is no visible change in the magnitudes of the relative error 
at $t\approx 3.5$~s.

The employed spatial resolution of $(N_r,N_z)=(32,64)$ grid points
is a common choice for real-time analysis in medium-sized tokamaks
\cite{lao85,lfg90,Barp20122112,gia2013,Moret20151}.
For the chosen AUG discharge a comparison with a run using a 
twofold finer spatial grid spacing, $(N_r,N_z)=(64,128)$, shows
good agreement for the majority of quantities, with
$q_{95}$ and $R_\mathrm{outer}$ exhibiting the largest sensitivity to
the resolution.

We conclude that by performing four iterations per time point and
using $(N_r,N_z)=(32,64)$ spatial grid points, sufficiently 
accurate equilibrium solutions for deriving 
real-time control parameters are obtained, at least in the sense of a
proof-of-principle using the chosen example  
data from AUG shot \#23221. 
A systematic analysis of the accuracy
and a comprehensive validation of the code
under various tokamak operational scenarios is beyond the scope of 
this paper.

\subsection{Computational Performance}\label{sect:performance}

\paragraph{Overview}

The computational performance of GPEC is assessed on a standard 
compute cluster with x86\_64 CPUs and InfiniBand (FDR~14) interconnect. 
The individual compute nodes are equipped with two Intel Xeon E5-2680v3 "Haswell" 
CPUs (2.5 GHz, 2x12 cores).
We use a standard Intel software stack (FORTRAN compiler v14.0, MPI v5.0) on top of the
Linux operating system (SLES11 SP3). For the required linear algebra
functionality the OpenBLAS \cite{openblas_url} library 
(version 0.2.13) is employed using the de-facto-standard interfaces from
BLAS \cite{BLAS_standard,netlib_url} and LAPACK
\cite{LAPACK,netlib_url}, and the FFTW \cite{FFTW05} library (version
3.3.4) is used for the 
discrete sine transforms in the Poisson solver. Both libraries are
open-source software released under the BSD or the GPL license, respectively. 
If desired, open-source alternatives to the Intel compilers (e.g. GCC \cite{gcc_url}) and 
MPI libraries (e.g. OpenMPI \cite{openmpi_url},
MVAPICH \cite{mvapich_url}) could be readily
utilized, albeit possibly with a certain performance impact.

Table~\ref{tab:pide_timing} shows an overview of the total 
runtime per time point and of the individual contributions from computing
the equilibrium solution and from deriving the set of plasma-control
parameters. We chose combinations of different numerical resolutions 
(in terms of the spatial grid resolution $N_r\times N_z$ and number of 
basis functions $N$) and computing resources (in terms of the number of
CPU cores).

With two compute nodes (each with 24 CPU cores) and using a
resolution of $(N_r,N_z,N\!+\!1)=(32,64,8)$ a runtime well below
1~ms and hence real-time capability is reached. When doubling the 
number of basis functions ($N\!+\!1\!=\!16$) real-time calculation is 
still possible on four compute nodes, indicating overall good weak 
scalability of the code. 

%
Roughly three quarters of the runtime is required for computing the 
equilibrium solution and the rest goes into evaluating the set of 
plasma control parameters. Hence, with a reduction of the number of
equilibrium iterations to only one or two
\cite{P-EFIT,Barp20122112,gia2013,Moret20151} runtimes of
0.5~ms are reached on just a single compute server.

\afterpage{\clearpage}

With a higher spatial resolution of $(N_r,N_z)=(64,128)$ a cycle time of 
2~ms is possible with 4 iterations, although from the linear
algorithmic complexity of the algorithm \cite{gpec12}, and the amount of data which
is communicated between the processes one would expect the runtime to
increase by a factor of four compared to $(N_r,N_z)=(32,64)$.
The reason is a higher parallel efficiency of the 
Poisson solver which is limited by OpenMP overhead at 
lower resolutions \cite{gpec12} as well as lower communication 
overhead due to larger message sizes in the expensive 
MPI\_Allreduce operation
which collects the distributed fields $\psi_k(r,z)$ and computes the 
sum $\psi = \sum_{k=1}^{N} c_k \psi_k$ 
(see the rows labelled "Poisson solver" and "$\psi$ sum" in Table~\ref{tab:psi_timing}). 


\begin{table*}[t]
\centering
\small
\begin{tabular}{|p{2.8cm}|lr|lr|lr|lr|}
\hline
resolution\newline $(N_r,N_z,N\!\!+\!\!1)$  &  
\multicolumn{2}{|c|}{$(32,64,8)$}&
\multicolumn{2}{|c|}{$(32,64,16)$}&
\multicolumn{2}{|c|}{$(64,128,8)$}&
\multicolumn{2}{|c|}{$(64,128,16)$}\\
\hline
cores (MPI tasks\newline $\cdot$ threads/task) &
\multicolumn{2}{|c|}{$48 (8\cdot 6)$}&
\multicolumn{2}{|c|}{$96 (16\cdot 6)$}&
\multicolumn{2}{|c|}{$48 (8\cdot 6)$}&
\multicolumn{2}{|c|}{$96 (16\cdot 6)$}\\
\hline
 &  $T$ [ms] &   \% &  $T$ [ms] &   \% &  $T$ [ms] &   \% &  $T$ [ms] &   \%   \\
\hline                       
RHS update          &  0.022  & 13.2    & 0.021 & 11.7  & 0.050  & 15.7   &  0.053   & 15.0\\
Poisson solver      &  0.036  & 21.6    & 0.036 & 19.5  & 0.085  & 26.6   &  0.086   & 24.5\\
RM evaluation       &  0.024  & 14.5    & 0.024 & 13.3  & 0.025  &  7.8   &  0.025   &  7.1\\
RM gather$^*$       &  0.015  &  8.9    & 0.021 & 11.4  & 0.019  &  5.9   &  0.029   &  8.1\\
lin.\ regression    &  0.013  &  7.9    & 0.017 &  9.5  & 0.014  &  4.4   &  0.018   &  5.2\\
$\psi$ sum$^{**}$     & 0.048  & 28.9    & 0.056 & 30.5  & 0.110  &  36.2  &  0.128  &  36.2 \\
other               &  0.008  & 5.0     & 0.008 &  4.2  & 0.011  &  3.4   &  0.014   & 3.9\\
\hline                                             
total               &  0.167  &  100.0  & 0.182 & 100.0 & 0.319   & 100.0  & 0.353 & 100.0 \\
\hline
\end{tabular}
\caption{Runtime breakdown of a single Picard equilibrium-iteration
  step (average over 4 iteration steps times 1000 time points) for
  different combinations (subset of Table~\ref{tab:pide_timing}) of numerical resolution and computational resources. The
  individual algorithmic steps from top to bottom correspond to the 
  sequence in Sect.~\ref{sect:algorithm_equil}. 
  Routines labelled with asterisks are dominated by MPI
  communication ($^*${\tt MPI\_Allgather}, $^{**}${\tt MPI\_Allreduce}).}
\label{tab:psi_timing}
\end{table*}

\paragraph{Equilibrium reconstruction}

Table~\ref{tab:psi_timing} further shows that the runtime for the equilibrium
iterations is dominated by computing the distributed sum, $\psi =
\sum_{k=1}^{N} c_k \psi_k$, which is also
an implicit synchronization point for all processes, and by 
the Poisson solver. Due to the use of high-order interpolation schemes
(cf.~Sect.\ref{sect:algorithm_equil}) a significant fraction
of about 25\% is spent on the evaluation of the
right-hand sides of  Eqs.~(\ref{gsphi1},\ref{gsphi2}) ("RHS update") and of the 
response matrix $\matr{B}$ ("RM evaluation"). 

In general, the chosen products of number of MPI
tasks times OpenMP threads per task turned out as the most efficient ones for
this hardware platform with 24 cores per node: using more than 6
threads (or 8 threads in the case of $(N_r,N_z)=(64,128)$) results
only in a modest speedup of
the Poisson solver \cite{gpec12} but doubles the required 
number of cores and network resources. Keeping those constant and
reducing the number of MPI tasks instead by a factor of two would require 
each MPI task to handle two basis functions. Although the
collective MPI communications would be somewhat faster in this case 
the resultant doubling of the effective runtime for the Poisson solver 
cannot be compensated for.

\paragraph{Plasma-control parameters}

\begin{figure*}[!ht]
  \centering
  \includegraphics[width=0.99\hsize,angle=0]{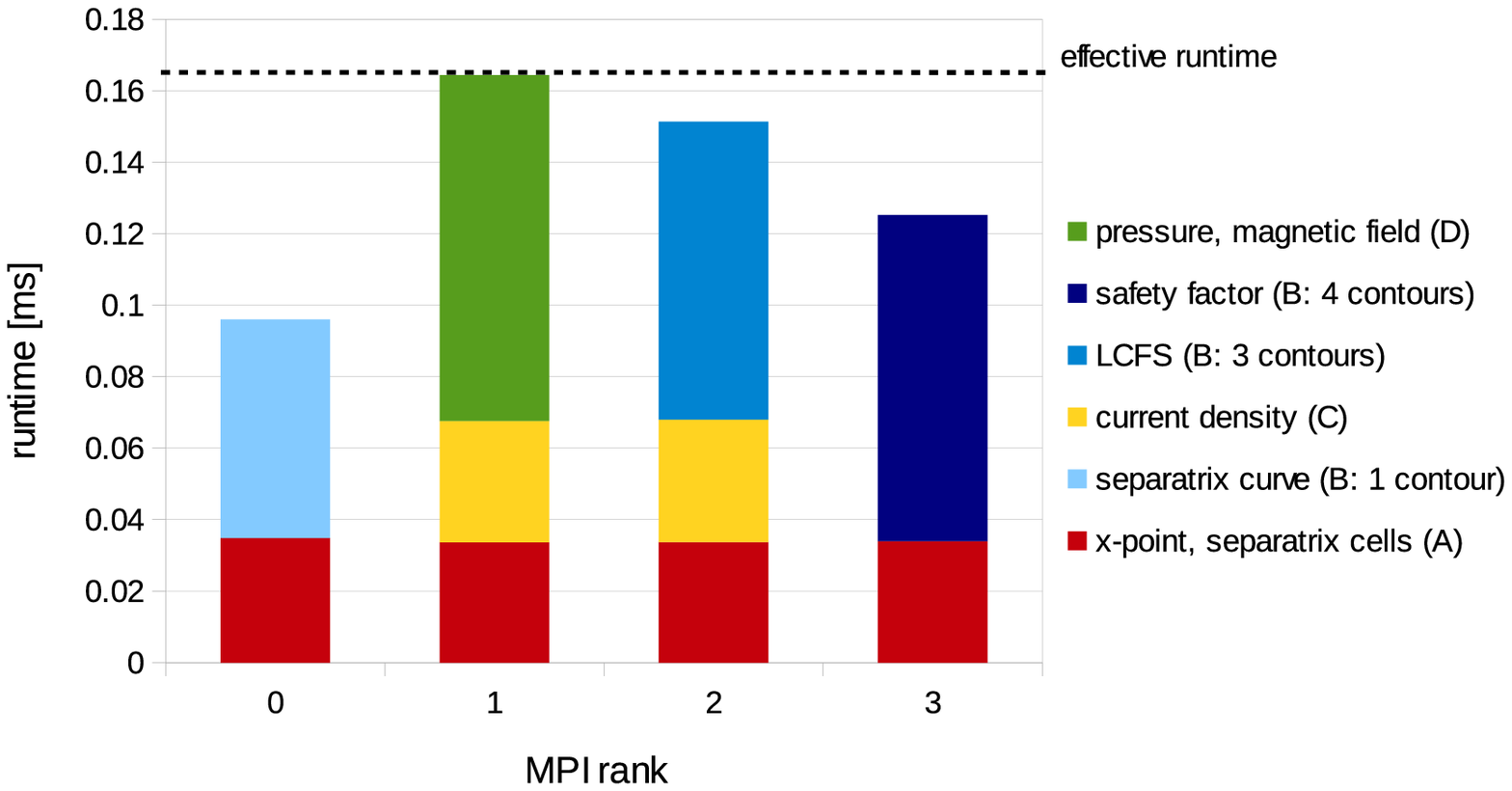}
  \caption{Runtime of the major groups (A-D) of computations for the
  plasma-control parameters (cf.\ Sect.~\ref{sect:postprocessing}) and their assignment to four  
  MPI tasks which execute in parallel. Within each task the
  individual parts of the algorithm (indicated by different colours) execute
  in the order from bottom to top. Accordingly, the dashed horizontal line
  marks the effective runtime in the application. 
  The data
  corresponds to AUG shot \#23221 computed with a resolution 
  of $(N_r,N_z,N\!+\!1)=(32,64,8)$ using in total 8 MPI tasks, each with 6 OpenMP threads (cf.\ Table~\ref{tab:pide_timing}).}
  \label{fig:pp_mpi}
\end{figure*}

For the example of $(N_r,N_z,N+1)=(32,64,8)$ computed with 48 CPU
cores, Figure~\ref{fig:pp_mpi} shows how the runtime of 
$T_\mathrm{ctl}=0.17$~ms (cf. Table~\ref{tab:pide_timing}) is 
composed and illustrates the benefits of exploiting task parallelism.
Due to the final {\tt MPI\_Allreduce} operation in the equilibrium
construction algorithm, which comes at little extra effective runtime
cost as compared to a single {\tt MPI\_Reduce}, all MPI tasks
hold a copy of the equilibrium flux distribution $\psi(r,z)$.
Hence, independent parts of the time-consuming computations for the 
plasma-control parameters like, e.g.\ evaluating the
pressure distribution or determining the contours for the safety
factor $q$, can be handled in parallel by different MPI tasks and
only scalar quantities need to be communicated (within a few
microseconds) between the MPI tasks at the end.
If necessary, the remaining load-imbalance indicated by Figure~\ref{fig:pp_mpi},
namely task 0 and task 3 becoming idle after about 0.1~ms, and tasks $4,\dotsc,7$ not computing
anything at all at this stage,
could be exploited for computing additional quantities and/or for 
further reducing the effective runtime.
Without task-parallelization, on the contrary, the runtimes would add
up to more than 0.4~ms despite the thorough OpenMP parallelization (e.g.\ across 
different contour levels) within the task.

\section{CONCLUSIONS}\label{sect:conclusions}

With the motivation of achieving sub-millisecond runtimes 
a new, parallel equilibrium-reconstruction code, GPEC, was presented
which is suitable for 
real-time applications in medium-sized tokamaks like ASDEX Upgrade (AUG).
GPEC implements the classical concept of
iteratively solving the Grad-Shafranov equation and feeding in diagnostic
signals from the experiment \cite{lfg90,cliste}.
Specifically, GPEC is implemented as a variant of the IDE code \cite{fis13}, which is 
a descendant of CLISTE \cite{cliste}. Compared
with these well-established and validated offline-analysis codes no algorithmic 
simplifications are necessary
for achieving the desired cycle times of less than a millisecond,
besides limiting the number of equilibrium iterations to four.
In addition to real-time applications 
the new code enables fast and highly accurate offline analyses soon 
after the tokamak discharge: Here, tolerable runtimes are in the range of
0.1--1~s per cycle which allows computing fully converged equilibria
employing highly resolved spatial and basis functions grids.

The parallelization of the new code builds on a fast shared-memory-parallel 
Grad-Shafranov solver \cite{gpec12} together with the MPI-distributed
solution of the individual Poisson-type problems and a thorough 
parallelization of the post-processing algorithms for
computing the relevant plasma-control parameters from the equilibrium flux distribution.

Using data from a typical AUG discharge the real-time 
capability of the new code was demonstrated by the offline computation of 
a sequence of 1000 time points
within less than a second of runtime. The relative accuracy was 
ascertained by
comparing the relevant plasma parameters with a converged run. 
By allowing four iterations for computing the equilibrium
solution the majority of control parameters can be computed with an
accuracy of a percent or better.

The adopted two-level, hybrid parallelization scheme allow
efficient utilization available compute resources (in terms of the
numbers of nodes, of CPU sockets per node and of cores per socket)
for a given numerical resolution (in terms of the spatial grid and
number of basis functions). 
Moreover, foreseeable advances in computer technology, in particular
the increasing number of CPU cores per compute node, are expected to push the limits
of real-time applications with GPEC towards even higher numerical accuracies 
(in terms of affordable resolution and number of equilibrium iterations).
We note that the benchmarks figures reported in this work can be
considered rather conservative as the computations
were performed on a standard compute cluster which comprises hundreds
of nodes. For energy-budget reasons such 
clusters are typically configured not with CPUs of the highest clock-frequency.
In our case, for example, CPUs with 2.5 GHz and only 24 cores per node were available.
For deployment in the control system of a tokamak experiment such as AUG, 
by contrast, we envision dedicated server hardware with higher clock 
frequencies and at least four CPU sockets, both of which is expected
to further boost the computational performance of the real-time
application. Thus it should be rather straightforward to save enough time
for the communication of the code with the control system, which,
depending of the specifics of the system, requires another 
few hundred microseconds per cycle.

Finally, it is worth mentioning that GPEC is based 
entirely on open-source software components, relies on established industry 
(FORTRAN, MPI, OpenMP) or de-facto software standards (BLAS, LAPACK, FFTW), runs on 
standard server hardware and software environments and hence can be 
released to the community and utilized without legal or commercial restrictions. 

\section*{ACKNOWLEDGEMENT}

We acknowledge stimulating discussions with K.\ Lackner and W.\ Treutterer.
Thanks to P.\ Martin who has provided a subversion of the CLISTE code with which
this study started and to S.\ Gori who has developed an initial FORTRAN~90-version.
\smallskip\\
This work has been carried out within the framework of the EUROfusion
   Consortium and has received funding from the Euratom research and
   training programme 2014-2018 under grant agreement No 633053. The views
   and opinions expressed herein do not necessarily reflect those of the
   European Commission.
\bibliographystyle{./fst}

\bibliography{./lit}

\begin{thebibliography}{10}

\bibitem{gormezano07}
\MakeUppercase{C.~Gormezano}, \MakeUppercase{A.~Sips}, \MakeUppercase{T.~Luce},
  \MakeUppercase{S.~Ide}, \MakeUppercase{A.~Becoulet},
  \MakeUppercase{X.~Litaudon}, \MakeUppercase{A.~Isayama},
  \MakeUppercase{J.~Hobirk}, \MakeUppercase{M.~Wade},
  \MakeUppercase{T.~Oikawa}, \MakeUppercase{R.~Prater},
  \MakeUppercase{A.~Zvonkov}, \MakeUppercase{B.~Lloyd},
  \MakeUppercase{T.~Suzuki}, \MakeUppercase{E.~Barbato},
  \MakeUppercase{P.~Bonoli}, \MakeUppercase{C.~Phillips},
  \MakeUppercase{V.~Vdovin}, \MakeUppercase{E.~Joffrin},
  \MakeUppercase{T.~Casper}, \MakeUppercase{J.~Ferron},
  \MakeUppercase{D.~Mazon}, \MakeUppercase{D.~Moreau},
  \MakeUppercase{R.~Bundy}, \MakeUppercase{C.~Kessel},
  \MakeUppercase{A.~Fukuyama}, \MakeUppercase{N.~Hayashi},
  \MakeUppercase{F.~Imbeaux}, \MakeUppercase{M.~Murakami},
  \MakeUppercase{A.~Polevoi}, \& \MakeUppercase{H.~S. John}.
\newblock Chapter 6: Steady state operation.
\newblock \emph{Nuclear Fusion}, \textbf{47}(6), S285 (2007).

\bibitem{gribov07}
\MakeUppercase{Y.~Gribov}, \MakeUppercase{D.~Humphreys},
  \MakeUppercase{K.~Kajiwara}, \MakeUppercase{E.~Lazarus},
  \MakeUppercase{J.~Lister}, \MakeUppercase{T.~Ozeki},
  \MakeUppercase{A.~Portone}, \MakeUppercase{M.~Shimada},
  \MakeUppercase{A.~Sips}, \& \MakeUppercase{J.~Wesley}.
\newblock Chapter 8: Plasma operation and control.
\newblock \emph{Nuclear Fusion}, \textbf{47}(6), S385 (2007).

\bibitem{mgg05}
\MakeUppercase{M.~Maraschek}, \MakeUppercase{G.~Gantenbein},
  \MakeUppercase{T.~P. Goodman}, \MakeUppercase{S.~G{\"u}nter},
  \MakeUppercase{D.~F. Howell}, \MakeUppercase{F.~Leuterer},
  \MakeUppercase{A.~M{\"u}ck}, \MakeUppercase{O.~Sauter}, \&
  \MakeUppercase{H.~Zohm}.
\newblock Active control of {MHD} instabilities by {ECCD} in {ASDEX} upgrade.
\newblock \emph{Nuclear Fusion}, \textbf{45}, 1369 (2005).

\bibitem{gia2013}
\MakeUppercase{L.~Giannone}, \MakeUppercase{M.~Reich},
  \MakeUppercase{M.~Maraschek}, \MakeUppercase{C.~Rapson},
  \MakeUppercase{W.~Treutterer}, \MakeUppercase{L.~Barrera},
  \MakeUppercase{E.~Poli}, \MakeUppercase{A.~Bock}, \MakeUppercase{G.~Conway},
  \MakeUppercase{F.~Fischer}, \MakeUppercase{J.~Fuchs},
  \MakeUppercase{K.~Lackner}, \MakeUppercase{P.~McCarthy},
  \MakeUppercase{A.~Mlynek}, \MakeUppercase{R.~Preuss}, \MakeUppercase{K.-H.
  Schuhbeck}, \MakeUppercase{J.~Stober}, \MakeUppercase{M.~Rampp},
  \MakeUppercase{R.~McDermott}, \MakeUppercase{Q.~Ruan},
  \MakeUppercase{L.~Wenzel}, \& \MakeUppercase{{ASDEX Upgrade Team}}.
\newblock A data acquisition system for real-time magnetic equilibrium
  reconstruction on {ASDEX Upgrade} and its application to {NTM} stabilization
  experiments.
\newblock \emph{Fusion Engineering and Design}, \textbf{88}, 3299 (2013).

\bibitem{sha57}
\MakeUppercase{V.~D. Shafranov}.
\newblock About {MHD} equilibrium configurations.
\newblock \emph{Zh. Eksp. Teor. Fiz.}, \textbf{33}, 710 (1957).

\bibitem{lus57}
\MakeUppercase{R.~L{\"u}st} \& \MakeUppercase{A.~Schl{\"u}ter}.
\newblock Axisymmetrische magnetohydrodynamische
  {Gleichgewichtskonfigurationen}.
\newblock \emph{Z. Naturforsch.}, \textbf{129}, 850 (1957).

\bibitem{lfg90}
\MakeUppercase{L.~L. Lao}, \MakeUppercase{J.~R. Ferron}, \MakeUppercase{R.~J.
  Groebner}, \MakeUppercase{W.~Howl}, \MakeUppercase{H.~{St. John}},
  \MakeUppercase{E.~J. Strait}, \& \MakeUppercase{T.~S. Taylor}.
\newblock Equilibrium analysis of current profiles in tokamaks.
\newblock \emph{Nuclear Fusion}, \textbf{30}, 1035 (1990).

\bibitem{cliste}
\MakeUppercase{P.~J. McCarthy}, \MakeUppercase{P.~Martin}, \&
  \MakeUppercase{W.~Schneider}.
\newblock The {CLISTE} interpretive equilibrium code.
\newblock Tech. Rep. IPP 5/85, Max-Planck-Institut f{\"u}r Plasmaphysik (1999).

\bibitem{lao85}
\MakeUppercase{L.~Lao}, \MakeUppercase{H.~S. John},
  \MakeUppercase{R.~Stambaugh}, \MakeUppercase{A.~Kellman}, \&
  \MakeUppercase{W.~Pfeiffer}.
\newblock Reconstruction of current profile parameters and plasma shapes in
  tokamaks.
\newblock \emph{Nuclear Fusion}, \textbf{25}(11), 1611 (1985).

\bibitem{blum12}
\MakeUppercase{J.~Blum}, \MakeUppercase{C.~Boulbe}, \&
  \MakeUppercase{B.~Faugeras}.
\newblock Reconstruction of the equilibrium of the plasma in a tokamak and
  identification of the current density profile in real time.
\newblock \emph{Journal of Computational Physics}, \textbf{231}(3), 960 (2012).

\bibitem{braams86}
\MakeUppercase{B.~Braams}, \MakeUppercase{W.~Jilge}, \&
  \MakeUppercase{K.~Lackner}.
\newblock Fast determination of plasma parameters through function
  parametrization.
\newblock \emph{Nuclear Fusion}, \textbf{26}(6), 699 (1986).

\bibitem{giannone10}
\MakeUppercase{L.~Giannone}, \MakeUppercase{M.~Cerna}, \MakeUppercase{R.~Cole},
  \MakeUppercase{M.~Fitzek}, \MakeUppercase{A.~Kallenbach},
  \MakeUppercase{K.~Lüddecke}, \MakeUppercase{P.~McCarthy},
  \MakeUppercase{A.~Scarabosio}, \MakeUppercase{W.~Schneider},
  \MakeUppercase{A.~Sips}, \MakeUppercase{W.~Treutterer},
  \MakeUppercase{A.~Vrancic}, \MakeUppercase{L.~Wenzel}, \MakeUppercase{H.~Yi},
  \MakeUppercase{K.~Behler}, \MakeUppercase{T.~Eich},
  \MakeUppercase{H.~Eixenberger}, \MakeUppercase{J.~Fuchs},
  \MakeUppercase{G.~Haas}, \MakeUppercase{G.~Lexa},
  \MakeUppercase{M.~Marquardt}, \MakeUppercase{A.~Mlynek},
  \MakeUppercase{G.~Neu}, \MakeUppercase{G.~Raupp}, \MakeUppercase{M.~Reich},
  \MakeUppercase{J.~Sachtleben}, \MakeUppercase{K.~Schuhbeck},
  \MakeUppercase{T.~Zehetbauer}, \MakeUppercase{S.~Concezzi},
  \MakeUppercase{T.~Debelle}, \MakeUppercase{B.~Marker},
  \MakeUppercase{M.~Munroe}, \MakeUppercase{N.~Petersen}, \&
  \MakeUppercase{D.~Schmidt}.
\newblock Data acquisition and real-time signal processing of plasma
  diagnostics on {ASDEX} upgrade using labview {RT}.
\newblock \emph{Fusion Engineering and Design}, \textbf{85}(3–4), 303
  (2010).
\newblock Proceedings of the 7th {IAEA} Technical Meeting on Control, Data
  Acquisition, and Remote Participation for Fusion Research.

\bibitem{P-EFIT}
\MakeUppercase{X.~N. Yue}, \MakeUppercase{B.~J. Xiao}, \MakeUppercase{Z.~P.
  Luo}, \& \MakeUppercase{Y.~Guo}.
\newblock Fast equilibrium reconstruction for tokamak discharge control based
  on {GPU}.
\newblock \emph{Plasma Physics and Controlled Fusion}, \textbf{55}(8), 085016
  (2013).

\bibitem{Barp20122112}
\MakeUppercase{A.~Barp}, \MakeUppercase{M.~Cerna}, \MakeUppercase{S.~Concezzi},
  \MakeUppercase{L.~Giannone}, \MakeUppercase{G.~Morrow},
  \MakeUppercase{Q.~Ruan}, \MakeUppercase{A.~Veeramani}, \&
  \MakeUppercase{L.~Wenzel}.
\newblock A real-time {Grad-Shafranov} {PDE} solver and {MIMO} controller.
\newblock \emph{Fusion Engineering and Design}, \textbf{87}(12), 2112  (2012).
\newblock Proceedings of the 8th {IAEA} Technical Meeting on Control, Data
  Acquisition, and Remote Participation for Fusion Research.

\bibitem{Moret20151}
\MakeUppercase{J.-M. Moret}, \MakeUppercase{B.~Duval}, \MakeUppercase{H.~Le},
  \MakeUppercase{S.~Coda}, \MakeUppercase{F.~Felici}, \&
  \MakeUppercase{H.~Reimerdes}.
\newblock Tokamak equilibrium reconstruction code {LIUQE} and its real time
  implementation.
\newblock \emph{Fusion Engineering and Design}, \textbf{91}, 1  (2015).

\bibitem{fer98}
\MakeUppercase{J.~Ferron}, \MakeUppercase{M.~Walker}, \MakeUppercase{L.~Lao},
  \MakeUppercase{H.~S. John}, \MakeUppercase{D.~Humphreys}, \&
  \MakeUppercase{J.~Leuer}.
\newblock Real time equilibrium reconstruction for tokamak discharge control.
\newblock \emph{Nuclear Fusion}, \textbf{38}, 1055 (1998).

\bibitem{yonekawa12}
\MakeUppercase{I.~Yonekawa}, \MakeUppercase{A.~V. Fernandez},
  \MakeUppercase{J.-M. Fourneron}, \MakeUppercase{J.-Y. Journeaux},
  \MakeUppercase{W.-D. Klotz}, \MakeUppercase{A.~Wallander}, \&
  \MakeUppercase{{CODAC Team}}.
\newblock {ITER} instrumentation and control system towards long pulse
  operation.
\newblock \emph{Plasma Fusion Research}, \textbf{7}, 2505047 (2012).

\bibitem{gpec12}
\MakeUppercase{M.~Rampp}, \MakeUppercase{R.~Preuss},
  \MakeUppercase{R.~Fischer}, \MakeUppercase{K.~Hallatschek}, \&
  \MakeUppercase{L.~Giannone}.
\newblock A parallel {Grad-Shafranov} solver for real-time control of tokamak
  plasmas.
\newblock \emph{Fusion Science and Technology}, \textbf{62}(3), 409 (2012).

\bibitem{fis13}
\MakeUppercase{R.~Fischer}, \MakeUppercase{J.~Hobirk}, \MakeUppercase{L.~B.
  Orte}, \MakeUppercase{A.~Bock}, \MakeUppercase{A.~Burckhart},
  \MakeUppercase{I.~Classen}, \MakeUppercase{M.~Dunne},
  \MakeUppercase{J.~Fuchs}, \MakeUppercase{L.~Giannone},
  \MakeUppercase{K.~Lackner}, \MakeUppercase{P.~McCarthy},
  \MakeUppercase{R.~Preuss}, \MakeUppercase{M.~Rampp},
  \MakeUppercase{S.~Rathgeber}, \MakeUppercase{M.~Reich},
  \MakeUppercase{B.~Sieglin}, \MakeUppercase{W.~Suttrop},
  \MakeUppercase{E.~Wolfrum}, \& \MakeUppercase{{ASDEX Upgrade Team}}.
\newblock Magnetic equilibrium reconstruction using geometric information from
  temperature measurements at asdex upgrade.
\newblock In \MakeUppercase{V.~Naulin}, \MakeUppercase{C.~Angioni},
  \MakeUppercase{M.~Borghesi}, \MakeUppercase{S.~Ratynskaia},
  \MakeUppercase{S.~Poedts}, \& \MakeUppercase{T.~Donné} (eds.), \emph{40th
  EPS Conference on Plasma Physics}, vol. 37D of \emph{Europhysics conference
  abstracts}, page P2.139. European Physical Society, Geneva (2013).

\bibitem{cliste12}
\MakeUppercase{P.~J. McCarthy} \& \MakeUppercase{{ASDEX Upgrade Team}}.
\newblock Identification of edge-localized moments of the current density
  profile in a tokamak equilibrium from external magnetic measurements.
\newblock \emph{Plasma Physics and Controlled Fusion}, \textbf{54}(1), 015010
  (2012).

\bibitem{hoe70}
\MakeUppercase{A.~E. Hoerl} \& \MakeUppercase{R.~W. Kennard}.
\newblock Ridge regression: Biased estimation for nonorthogonal problems.
\newblock \emph{Technometrics}, \textbf{12}(1), 55 (1970).

\bibitem{pre07}
\MakeUppercase{W.~H. Press}, \MakeUppercase{S.~A. Teukolsky},
  \MakeUppercase{W.~T. Vetterling}, \& \MakeUppercase{B.~P. Flannery}.
\newblock \emph{Numerical Recipes: The Art of Scientific Computing}.
\newblock Cambridge University Press, 3rd edn. (2007).

\bibitem{hal75}
\MakeUppercase{K.~v.~Hagenow} \& \MakeUppercase{K.~Lackner}.
\newblock Computation of axisymmetric {MHD} equilibria.
\newblock In \emph{Proc. 7th Conf. on the Numerical Simulation of Plasmas},
  page 140. New York, US (1975).

\bibitem{lac76}
\MakeUppercase{K.~Lackner}.
\newblock Computation of ideal {MHD} equilibria.
\newblock \emph{Comp. Phys. Comm.}, \textbf{12}, 33 (1976).

\bibitem{gpec}
\MakeUppercase{R.~Preuss}, \MakeUppercase{R.~Fischer},
  \MakeUppercase{M.~Rampp}, \MakeUppercase{K.~Hallatschek},
  \MakeUppercase{U.~von Toussaint}, \MakeUppercase{L.~Giannone}, \&
  \MakeUppercase{P.~McCarthy}.
\newblock Parallel equilibrium algorithm for real-time control of tokamak
  plasmas.
\newblock Tech. Rep. IPP R/47, Max-Planck-Institut f{\"u}r Plasmaphysik (2012).

\bibitem{mpiforum_url}
http://www.mpi-forum.org/.

\bibitem{stacey12}
\MakeUppercase{W.~M. Stacey}.
\newblock \emph{Fusion Plasma Physics}.
\newblock Wiley, 2 edn. (2012).

\bibitem{openblas_url}
http://www.openblas.net.

\bibitem{BLAS_standard}
\MakeUppercase{{BLAST Forum}}.
\newblock Basic linear algebra subprograms technical forum standard.
\newblock \emph{International Journal of High Performance Applications and
  Supercomputing}, \textbf{16}(1) (2002).

\bibitem{netlib_url}
http://www.netlib.org.

\bibitem{LAPACK}
\MakeUppercase{E.~Anderson}, \MakeUppercase{Z.~Bai},
  \MakeUppercase{C.~Bischof}, \MakeUppercase{S.~Blackford},
  \MakeUppercase{J.~Demmel}, \MakeUppercase{J.~Dongarra},
  \MakeUppercase{J.~Du~Croz}, \MakeUppercase{A.~Greenbaum},
  \MakeUppercase{S.~Hammarling}, \MakeUppercase{A.~McKenney}, \&
  \MakeUppercase{D.~Sorensen}.
\newblock \emph{{LAPACK} Users' Guide}.
\newblock Society for Industrial and Applied Mathematics, Philadelphia, PA,
  third edn. (1999).
\newblock ISBN 0-89871-447-8 (paperback).

\bibitem{FFTW05}
\MakeUppercase{M.~Frigo} \& \MakeUppercase{S.~G. Johnson}.
\newblock The design and implementation of {FFTW3}.
\newblock \emph{Proceedings of the IEEE}, \textbf{93}(2), 216 (2005).
\newblock Special issue on ``Program Generation, Optimization, and Platform
  Adaptation''.

\bibitem{gcc_url}
https://gcc.gnu.org/.

\bibitem{openmpi_url}
http://www.open-mpi.de/.

\bibitem{mvapich_url}
http://mvapich.cse.ohio-state.edu/.

\end{thebibliography}

\end{document}